\documentclass[reprint,aps,pre]{revtex4-1}
\usepackage{amsmath,graphicx,multirow,array}
\usepackage{amsthm,amssymb}
\usepackage{CJK}
\usepackage{upgreek}
\usepackage{subfig}

\usepackage[usenames,dvipsnames]{color}
\usepackage[normalem]{ulem}

\begin{document}
\begin{CJK*}{UTF8}{}

\title{Binding potentials for vapour nanobubbles on surfaces using density functional theory}

\author{Hanyu Yin (\CJKfamily{gbsn}尹寒玉)}
\affiliation{Department of Mathematical Sciences,
Loughborough University, Loughborough, LE11 3TU, UK}

\author{David N. Sibley}
\affiliation{Department of Mathematical Sciences,
Loughborough University, Loughborough, LE11 3TU, UK}

\author{Andrew J. Archer}
\affiliation{Department of Mathematical Sciences,
Loughborough University, Loughborough, LE11 3TU, UK}

\date{\today}

\begin{abstract}
We calculate density profiles of a simple model fluid in contact with a planar surface using density functional theory (DFT), in particular for the case where there is a vapour layer intruding between the wall and the bulk liquid. We apply the method of Hughes \emph{et al.}\ [J.\ Chem.\ Phys.\ {\bf 142}, 074702 (2015)] to calculate the density profiles for varying (specified) amounts of the vapour adsorbed at the wall. This is equivalent to varying the thickness $h$ of the vapour at the surface. From the resulting sequence of density profiles we calculate the thermodynamic grand potential as $h$ is varied and thereby determine the binding potential as a function of $h$. The binding potential obtained via this coarse-graining approach allows us to determine the disjoining pressure in the film and also to predict the shape of vapour nano-bubbles on the surface. Our microscopic DFT based approach captures information from length scales much smaller than some commonly used models in continuum mechanics.
\end{abstract}

\maketitle

\end{CJK*}

\section{Introduction}
For more than two decades there has been interest in surface nanobubbles, which can form when a hydrophobic surface is fully immersed in liquid {\cite{parker1994bubbles, craig2011very, lohse2015surface, alheshibri2016history}}. Due to the high Laplace pressure inside a hemispherical cap shaped nanobubble, we might expect the gas inside to dissolve and diffuse away in microseconds \cite{ljunggren1997lifetime}. However, in reality they can sometimes remain stable for many hours or even up to days \cite{craig2011very, lohse2015surface, stevens2005effects, simonsen2004nanobubbles}. The existence of surface nanobubbles at the solid-liquid interface plays a significant role in a number of chemical and physical processes, such as flotation in mineral processing \cite{hampton2009accumulation}, design of microdevices \cite{paxton2004catalytic} and drug delivery to cancer cells \cite{janib2010imaging}. As well as the wide range of applications, there are also theoretical challenges to understanding the fundamental physical properties of nanobubbles which has also attracted the attention of many scientists. These surface nanobubbles contain air molecules that have come out of solution in the liquid, and are not purely filled with the vapour phase. To properly describe such a system, one must treat the full two component system of solvent liquid and solute air molecules. However, as a precursor to tackling the full binary mixture problem, the situation that must be first understood is that of the pure liquid and the properties of nanobubbles of the vapour that may appear between the liquid and a solid surface. It is this aspect that we discuss in the present paper. Our approach is to use a microscopic (i.e.\ particle resolved) classical density functional theory (DFT) \cite{evans1979nature, hansen2013theory} based method to calculate a coarse grained effective interfacial free energy (often called the binding potential, which is defined below) for vapour nanobubbles. { There are, of course, other computer simulation methods by which this can be done \cite{macdowell2011computer, tretyakov2013parameter, benet2016premelting, jain2019using}. The resulting binding potential} is then input into a mesoscopic interfacial free energy functional for determining the height profile of the nanobubbles. This also allows us to calculate the total free energy of such a nanobubble and how it depends on the interaction potential between the surface and the fluid particles, thereby allowing us to estimate the relative probabilities for observing nanobubbles as a function of size and surface properties.

\begin{figure}[b]
\includegraphics[width=1.\columnwidth]{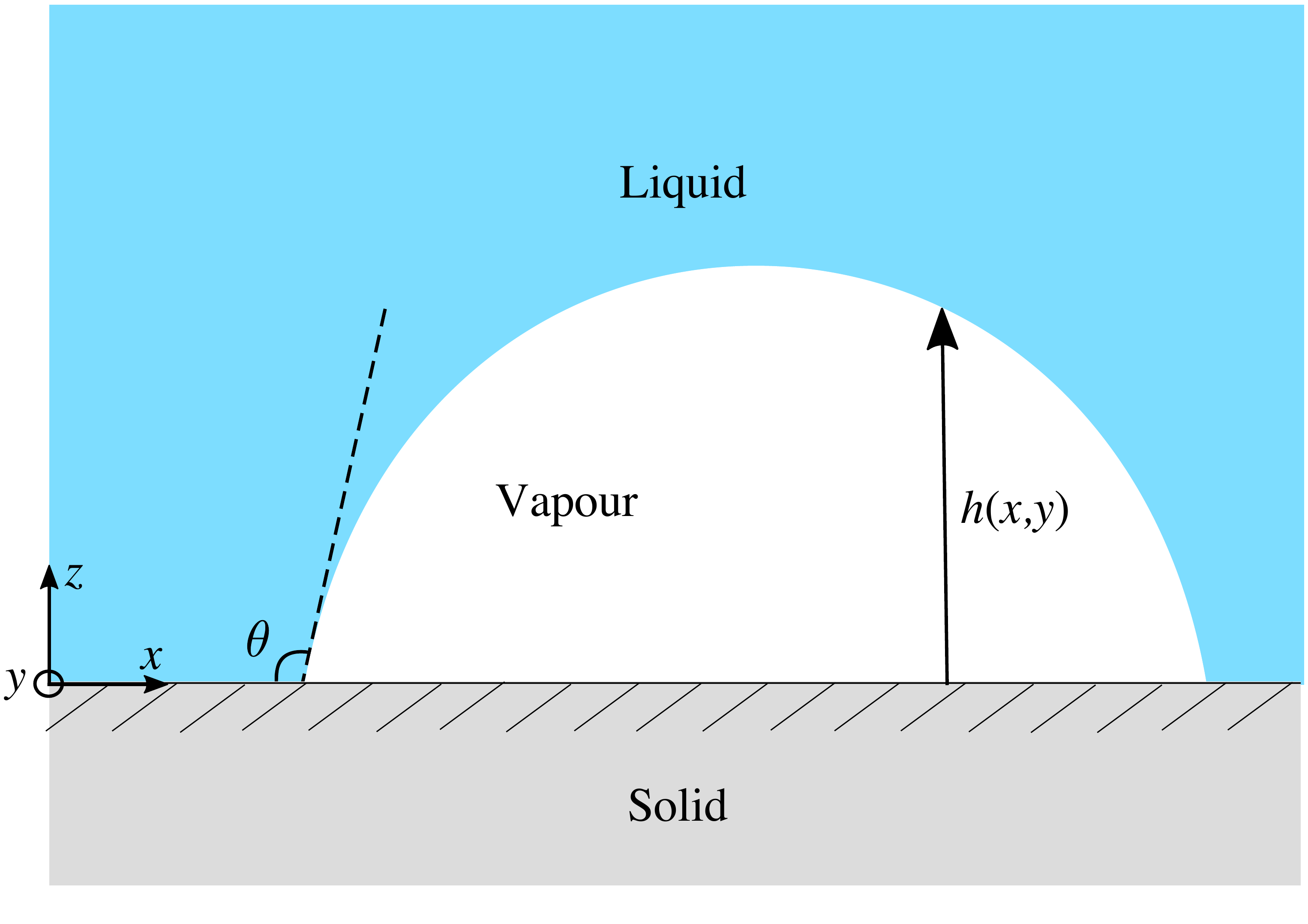}
\caption{\label{2dbubble} Sketch of a vapour bubble with height profile $z=h(x,y)$, surrounded by liquid, on top of a solid planar wall that exerts an external potential $V_{ext}(z)$ on the fluid. The coordinate direction $z$ is perpendicular to the solid surface and the $x$- and $y$-axes are parallel to the surface. {The contact angle of the liquid with the wall is $\theta$.}}
\end{figure}

The system we model here is a very small bubble of vapour located on a planar solid surface that is in contact with a bulk liquid. The height of the liquid-vapour interface is defined to be at $h(x,y)$ above the surface, where $(x,y)$ is the position on the surface. A sketch of the system is displayed in Fig.\ \ref{2dbubble}, illustrating a cross section through a (nanometre scaled) vapour bubble. To develop an understanding of such a bubble, $h(x,y)$ is a key quantity to be determined, as is the contact angle the liquid-vapour interface makes with the substrate. This, via Young's equation \cite{de2013capillarity}, is related to thermodynamic quantities, namely the three interfacial tensions: $\gamma_{lv}$, $\gamma_{sl}$ and $\gamma_{sv}$, which are the liquid-vapour, solid-liquid and solid-vapour interfacial tensions, respectively. Of course, for larger bubbles $h(x,y)$ is of the shape of a hemispherical cap, because this minimises the area of the liquid-vapour interface and so also the free energy of the system. However, near the contact line { (i.e.\ where the three phases meet)} there is an additional contribution to the free energy from the binding (or interfacial) potential $g(h)$, which results from molecular interactions. This influences the shape of $h(x,y)$ near the contact line and for nanobubbles is particularly important and can influence the overall shape of $h(x,y)$. The contribution to the pressure within the bubble can be expressed in terms of the Derjaguin (or disjoining) pressure $\Pi(h) = - \partial g(h)/\partial h$ \cite{de2013capillarity} and its effects can be observed experimentally \cite{zhang2008thermodynamic}.

The physics of vapour bubbles on surfaces shares many similarities with the more commonly studied system of liquid droplets on a surface, surrounded by the vapour. In both cases, the two main contributions to the excess free energy $F[h]$ of the system due to the interface are the binding potential contribution (i.e.\ due to the molecular interactions), and the surface tension contribution (proportional to the area of the liquid-vapour interface), which gives \cite{dietrich1988inphase, schick1990liquids, de2013capillarity}
\begin{equation}\label{IH}
F_{\textrm{IH}}\left[h\right] = \iint\left[g\left(h\right)+\gamma_{lv}\sqrt{1+\left(\nabla h\right)^2}\right]\mathrm dx\mathrm dy.
\end{equation}
This free energy is often termed an interfacial Hamiltonian (IH). Note that in Eq.~(\ref{IH}) we have omitted terms independent of $h$ -- see Eq.~(\ref{realbp}) below.

To study nanobubbles, in Ref.~\cite{svetovoy2016effect} a simple approximate form for the binding potential $g(h)$ was postulated, since although much can be inferred about the qualitative form of $g(h)$ from various considerations \cite{dietrich1988inphase, schick1990liquids, de2013capillarity}, its precise form is not known exactly. The model of Ref.~\cite{svetovoy2016effect} includes contributions to $g(h)$ due to the van der Waals forces. Our approach here is to develop a model for vapour nanobubbles at equilibrium, based on calculating the binding potential $g(h)$ using DFT for all values of $h$, that can then be used as an input to the IH model. Since DFT incorporates the effects of the compressibility of the vapour, these effects are also incorporated into $g(h)$ when it is calculated using our approach.

DFT is a hugely powerful and widely used microscopic statistical mechanical theory for calculating the density profile $\rho(\mathbf{r})$ for inhomogeneous systems of interacting particles, where $\mathbf{r}=(x,y,z)$. An advantage of DFT is that it gives a molecular-level detail description (as does, e.g.\ molecular dynamics computer simulations), but the computer time taken to solve DFT is small, particularly when the fluid average density profile only varies in one direction (e.g.\ perpendicular to the wall). DFT is especially suitable for determining excess thermodynamic quantities, arising from inhomogeneities in the fluid density distribution due to the presence of interfaces. There are numerous works applying DFT to study the wetting and drying interfacial phase behaviour of liquids -- see for example Refs.~\cite{meister1985density, tarazona1987phase, dietrich1988inphase, van1991wetting, henderson1992weighted, evans92, wu2006density, hughes2014introduction, evans2017drying, andreasPoFstructure, LuisPaper}. { Since DFT is an accurate theory for the spatial variations in the particle density, it thereby incorporates the effects of vapour compressibility, which are believed to be important for nanobubbles.} 

To determine the binding potential using DFT, one must calculate a series of constrained density profiles, the constraint being that the adsorption { $\Gamma$ (rather than the vapour thickness $h$) takes a series of specified values. Recall that $\Gamma=N_{ex}/A$, where $N_{ex}$ is the excess number of particles in the system due to the presence of the interface, which has area $A$ \cite{evans1990fluids}. Constraining $\Gamma$} can be done using the method proposed in Ref.~\cite{archer2011nucleation} and further developed by Hughes {\it et al.}\ \cite{hughes2015liquid, hughes2017influence}. These works showed that the required constraint takes the form of a {\em fictitious} external potential that can be calculated self-consistently as part of the algorithm for determining the constrained density profile. Hughes {\it et al.}\ \cite{hughes2015liquid, hughes2017influence} applied the method to determine the binding potential for films of liquid adsorbed on a surface in contact with a bulk vapour. Taking the resulting binding potentials together with the IH \eqref{IH} results in droplet profiles that are in excellent agreement with those obtained from solving the full DFT to determine the droplet profile \cite{hughes2015liquid, hughes2017influence}, validating the overall coarse graining approach. Further validation comes from Ref.~\cite{buller2017nudged} where two other completely different approaches for obtaining $g(\Gamma)$ were used that nonetheless produce identical results. These two approaches are: (i) applying the nudged-elastic-band algorithm to connect the sequence of density profiles required to calculate $g$ and (ii) a method based on an overdamped nonconserved dynamics to explore the underlying free-energy landscape. For liquid droplets, the resulting binding potential can also be input into a thin film hydrodynamic equation to study the dynamics of liquid droplets on surfaces \cite{yin2017films}.

Having calculated the binding potential $g$ as a function of the adsorption $\Gamma$, it is straightforward to relate this to the height $h$ of the vapour-liquid interface above the surface of the substrate. Note, however, that the adsorption $\Gamma$ is a more appropriate measure of the amount of a particular phase on a substrate than the height $h$ when the amounts are small and on microscopic length scales, e.g.\ when there is sub-monolayer adsorption at an interface \cite{hughes2015liquid, hughes2017influence,yin2017films}. The adsorption is defined as
\begin{equation}
\Gamma (x,y)= \int_0^\infty(\rho(\mathbf r) - \rho_b)  \mathrm {d}z,
\label{eq:ads}
\end{equation}
 where $\rho_b$ is the bulk fluid density and we have assumed the $z$-axis is perpendicular to the substrate, which has its planar surface at $z=0$. The corresponding height $h(x,y)$, quantifying the amount of the phase that is on the substrate, may be defined in a number of ways. This lack of a unique definition is another reason why $\Gamma(x,y)$ is a better measure. For example, one could define $h(x,y)$ to be the position where the average density $\rho(x,y,z=h)=(\rho_b+\rho_a)/2$, which is the average of the bulk density and the density of the phase adsorbed on the substrate, $\rho_a$. However, here we prefer to define $h$ as \cite{hughes2015liquid, hughes2017influence, yin2017films}
\begin{equation}
h(x,y) \equiv  \frac{\Gamma(x,y)}{\rho_a-\rho_b}.
\label{eq:bubble_height}
\end{equation}
In the situation where the bulk phase is the vapour (with density $\rho_b=\rho_v$) and the phase adsorbed on the surface is the liquid (with density $\rho_a=\rho_l$), then this is a widely used definition. Note also that in the case when the liquid is the bulk phase ($\rho_b=\rho_l$) and it is the vapour that is adsorbed at the interface ($\rho_b=\rho_v$), then in general both of the quantities in the numerator and denominator on the right hand side of Eq.~\eqref{eq:bubble_height} are negative, but of course still giving a positive thickness $h$.

This paper is structured as follows: Some background on the relevant interfacial thermodynamics and the definition of $g(h)$ is given in Sec.~\ref{sec:int_thermo}. In Sec.~\ref{sec:DFT_approach} we describe briefly the DFT based method we apply for calculating $g(h)$ for vapour films adsorbed between a planar wall and a bulk liquid. Then, in Sec.~\ref{sec:model_fluid}, we introduce the model fluid that we consider, the approximate DFT used to treat this fluid and the various different wall potentials that we consider. In Sec.~\ref{sec:results} we present results for $g(\Gamma)$, for various different wall potentials and how the decay form of the wall potential moving away from the wall influences the decay form of $g(\Gamma)$. Following this, in Sec.~\ref{sec:bubble_profiles} we input the obtained binding potentials into the interfacial Hamiltonian \eqref{IH}, in order to determine vapour nanobubble height profiles and their free energies. Finally, in Sec.~\ref{sec:conc} we draw our conclusions.

\section{Interfacial thermodynamics for vapour adsorption}\label{sec:int_thermo}

 \begin{figure}[t]
\includegraphics[width=1.\columnwidth]{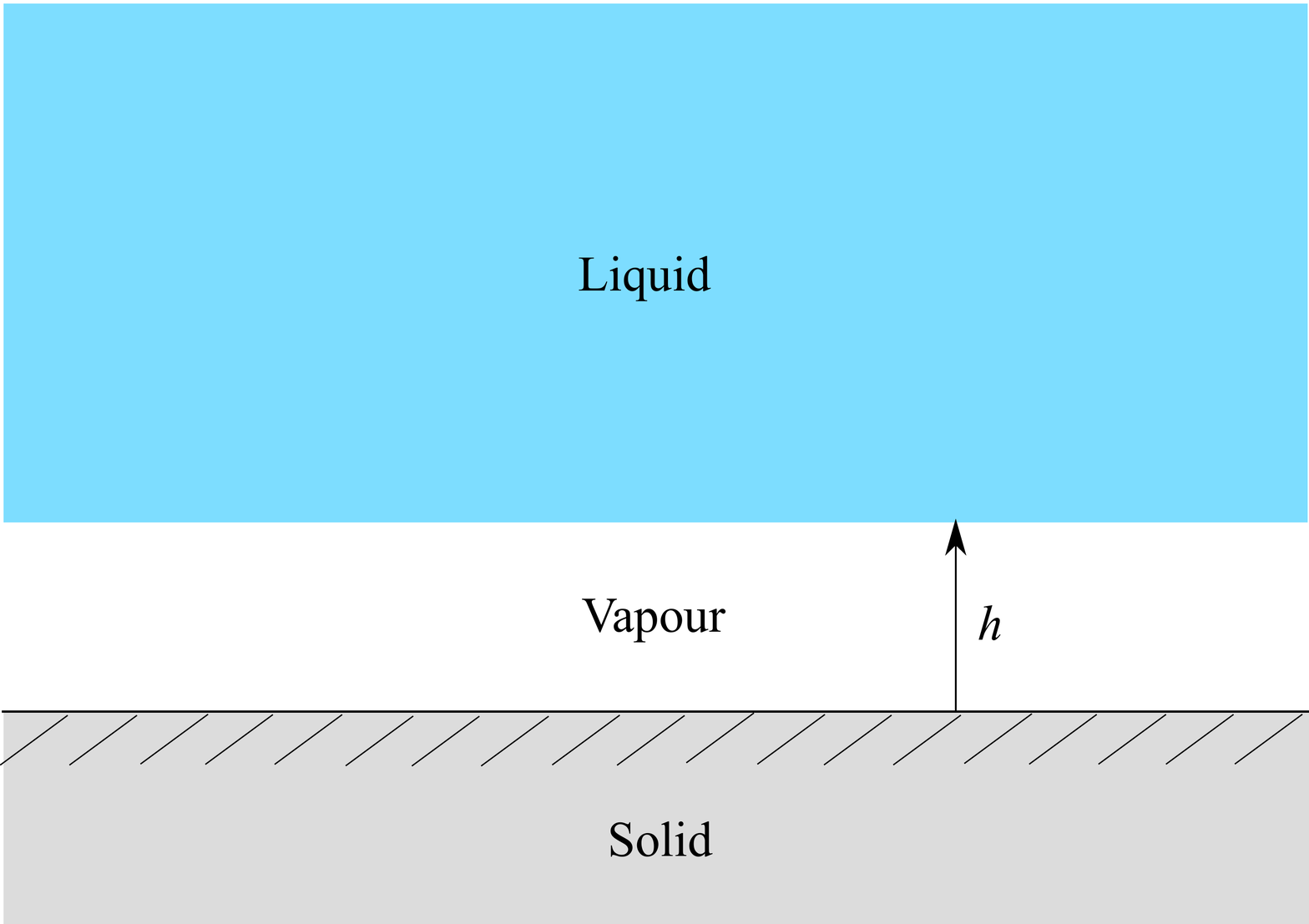}
\caption{\label{1dinterface}A schematic diagram of a uniform thickness layer of vapour adsorbed at the interface between a planar solid substrate and the bulk liquid. The thickness of the vapour film is $h$.}
\end{figure}

Consider the system illustrated in Fig.~\ref{1dinterface}. Treating it in the grand canonical ensemble, the grand potential $\Omega$ is the relevant free energy to consider, which is minimised when the system is at equilibrium. To describe the interfacial phase behaviour, we follow the usual procedure \cite{rowlinson1982molecular} and consider surface excess quantities; in this case it is the excess grand potential per unit area
\begin{equation}
\frac{\Omega_{ex}}{A} = \frac{\Omega - \Omega_b}{A},
\end{equation}
where $\Omega_b=-pV$ is the grand potential for a bulk system having the same volume $V$ and pressure $p$, but with no interface and where $A$ is the area of the wall. This can be split into the following contributions
\begin{equation}\label{excessbubble}
\frac{\Omega_{ex}(h)}{A} = \gamma_{lv}+\gamma_{sv} + h\delta p+g(h),
\end{equation}
where $\delta p=p-p_v$ is the pressure difference between the pressure of the bulk liquid and that of the corresponding vapour at the same chemical potential $\mu$. If the system is at bulk vapour-liquid coexistence, then this term is zero. The interfacial tensions $\gamma_{lv}$ and $\gamma_{sv}$ can be calculated using DFT in the usual way \cite{evans1979nature, hansen2013theory, wu2006density, hughes2014introduction}. The above equation may be viewed as defining the binding potential: it is the `remainder' after the other terms have been subtracted, i.e.\ at bulk vapour-liquid coexistence, with $\delta p = 0$, we have \cite{macdowell2011computer}
\begin{equation}\label{realbp}
g(\Gamma) = \frac{\Omega+pV}{A} - \gamma_{lv}-\gamma_{sv}.
\end{equation}
When $\Gamma$ $\rightarrow \infty$ the two interfaces are far from one another, so they do not influence each other, and therefore we have $g(\Gamma)\rightarrow 0$. However, when $\Gamma=\Gamma_0$, the value at the minimum of the binding potential, we have \cite{de2013capillarity}
\begin{equation}
g(\Gamma_0) = \gamma_{sl}-\gamma_{sv}-\gamma_{lv}.
\end{equation}
Using Young equation \cite{young1805essay} $\gamma_{lv}\cos\theta = \gamma_{sv}-\gamma_{sl}$, we obtain \cite{de2013capillarity, rauscher2008wetting, DeCh1974jcis}
\begin{equation}\label{youngs}
\cos \theta = \frac{\gamma_{sv}-\gamma_{sl}}{\gamma_{lv}}=-1-\frac{g(\Gamma_0)}{\gamma_{lv}},
\end{equation}
where $\theta$ is the equilibrium contact angle, measured as in the usual definition as the angle through {the} liquid phase. Therefore, this is the outer angle on bubbles and so we have the opposite sign in this equation compared to when considering liquid drops.

{ Note that if the system is away from coexistence, with $\delta p\neq0$, then the equilibrium state is not at $\Gamma=\Gamma_0$, the value at the minimum of $g(\Gamma)$. Instead, by minimising the excess grand potential in Eq.~\eqref{excessbubble} with respect to variations in $h$, we see that the equilibrium is given by $\frac{\partial}{\partial h}(h\delta p+ g(h))=0$, i.e.\ the equilibrium film thickness is the solution of $g'(h)+\delta p=0$. When $\delta p$ is small it can also be useful to use the Gibbs-Duhem relation $(\partial p/\partial \mu)_T=\rho$ (see e.g.\ \cite{evans1987phase}) to show that $\delta p=\Delta\rho\delta\mu$ when $\delta\mu$ is small, where $\Delta\rho=(\rho_l-\rho_v)$ and $\delta\mu=(\mu-\mu_{coex})$, enabling one to determine the equilibrium film thickness (i.e.\ adsorption) as a function of $\delta\mu$.}

\section{DFT approach to calculate $g(\Gamma)$}\label{sec:DFT_approach}

In DFT \cite{evans1979nature, hansen2013theory} we find that the grand potential is the following functional of the fluid density profile $\rho(\mathbf{r})$:
\begin{equation}\label{grand}
\Omega[\rho(\mathbf{r})] = F[\rho(\mathbf{r})] + \int \rho(\mathbf{r})(V_{ext}(\mathbf{r})-\mu) \mathrm{d}\mathbf{r},
\end{equation}
where $V_{ext}(\mathbf{r})$ is the external potential felt by a single particle at position $\mathbf{r}$ (i.e.\ the potential due to the solid substrate in the treatment here), $\mu$ is the chemical potential and
\begin{equation}\label{Fe}
F[\rho(\mathbf{r})]=k_BT\int\rho(\mathbf{r})(\ln[\Lambda^3\rho(\mathbf{r})]-1)\mathrm{d}\mathbf{r} +F_{ex}[\rho(\mathbf{r})]
\end{equation}
is the intrinsic Helmholtz free energy. The first term is the ideal-gas contribution and $F_{ex}$ is the excess part due to the interactions between the fluid particles. In the ideal-gas part, $k_B$ is Boltzmann's constant, $T$ is the temperature and $\Lambda$ is the thermal de Broglie wavelength. The equilibrium fluid density profile is that which minimises $\Omega[\rho(\mathbf{r})] $, i.e.\ it satisfies the Euler-Lagrange equation
\begin{equation}\label{euler}
\frac{\delta \Omega}{\delta \rho(\mathbf r)}=k_BT\ln[\Lambda^3\rho(\mathbf{r})]+\frac{\delta F_{ex}}{\delta\rho}+V_{ext}(\mathbf{r})-\mu=0.
\end{equation}
This equation may be rearranged to obtain
\begin{equation}
\rho(\mathbf{r})=\Lambda^{-3}e^{\beta[\mu-\frac{\delta F_{ex}}{\delta\rho}-V_{ext}(\mathbf{r})]},
\label{density}
\end{equation}
where $\beta=(k_BT)^{-1}$.
This is the form usually used for solving DFT numerically using a Picard iterative process \cite{hughes2014introduction, roth2010}. This consists of constructing a sequence of approximate solutions, indexed by the integer $k$, such that the $(k+1)$th approximation is obtained from the previous $k$th approximation, and with each successively closer to the true density profile. We start by guessing an initial density profile (for example the ideal-gas result), and calculate a new profile $\rho_{rhs}$ via the right hand side of Eq.~(\ref{density}). Then, a fraction of this new profile is mixed with the previous approximation for the profile $\rho_{k}$, to compute the new approximation 
\begin{equation}
\rho_{k+1}=\alpha\rho_{rhs}+(1-\alpha)\rho_{k}.
\label{eq:mixing}
\end{equation} 
This equation is then iterated till convergence to the desired tolerance is achieved. Here, $\alpha$ is the mixing parameter, which is typically in the range $0.1>\alpha>0.01$, for the algorithm to be numerically stable.

Solving the Euler-Lagrange equation (\ref{euler}) as described above gives the equilibrium fluid density profile that has adsorption $\Gamma_0$, as determined by Eq.~\eqref{eq:ads}. On substituting this density profile into Eq.~(\ref{grand}), together with Eq.~\eqref{realbp}, we obtain the minimum value of the binding potential. To find the full binding potential curve $g(\Gamma)$, requires calculating for a series of points over a range of different values of the adsorption $\Gamma$. As mentioned above, we do this by applying the fictitious potential approach developed and applied in Refs.\ \cite{archer2011nucleation, hughes2015liquid, hughes2017influence, buller2017nudged}. This method constrains the adsorption of the system to be a desired value by modifying the Picard iteration by replacing $\rho_{rhs}$ in Eq.~\eqref{eq:mixing} with
\begin{equation}
\rho_{new}=(\rho_{rhs}-\rho_b)\frac{\Gamma_d}{\Gamma_{rhs}}+\rho_b,
\end{equation}
where $\Gamma_{rhs}$ is the adsorption corresponding to the profile $\rho_{rhs}$ calculated via Eq.~\eqref{eq:ads} and $\Gamma_d$ is the desired value of the adsorption.             

\begin{figure}[t]
\includegraphics[width = 1.\columnwidth]{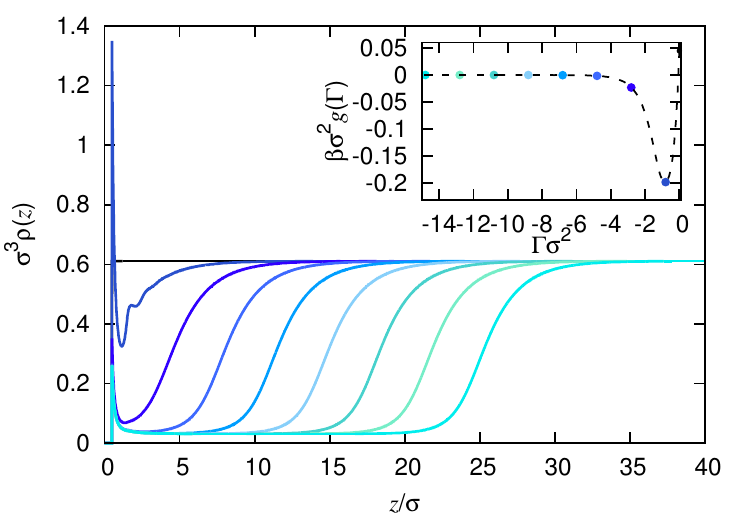}
\caption{\label{density_09}A sequence of density profiles with decreasing adsorption corresponding to increasing thickness films of vapour between a wall and the bulk liquid. The adsorption values for each are $\Gamma\sigma^2=-0.0$, $-0.8$, $-2.8$, $-4.8$, $-6.8$, $-8.8$, $-10.8$, $-12.8$ and $-14.8$, { where $\sigma$ is the diameter of the cores of the particles}. The strength of the attraction between the fluid particles is $\beta\epsilon=0.5$, with { range $\lambda=\sigma$}, and the system is at vapour-liquid coexistence, with $\mu=\mu_{coex}$. The wall potential is that in Eq.~\eqref{Yukawall}, with $\beta\epsilon_{w}^{(Y)}=1.817$ and $\lambda_{w}^{(Y)}/\sigma=1$. The inset shows the resulting binding potential, with the points on the curve corresponding to the sequence of density profiles displayed in the main figure.}         
\end{figure}

A typical series of the constrained density profiles calculated using this procedure are displayed in Fig.~\ref{density_09}. These results are for the model fluid defined below, with fixed wall attraction strength. The inset shows the corresponding binding potential $g(\Gamma)$. The global minimum occurs at a small negative value of the adsorption, which corresponds to a partially drying liquid. In the density profiles there is peak near to the wall, corresponding to some particles being adsorbed preferentially at a particular distance from the surface of the wall. In the second density profile, which corresponds to the minimum in the binding potential, there are some oscillations near the wall, due to packing effects of the particles. As the adsorption becomes increasingly negative, there is an increasingly thick film of the vapour near the wall, and also as the thickness increases, the vapour density in the film becomes closer to that of the vapour at bulk vapour-liquid coexistence.

\section{Model fluid}\label{sec:model_fluid}

\begin{figure}[t]
 \includegraphics[width=1.\columnwidth]{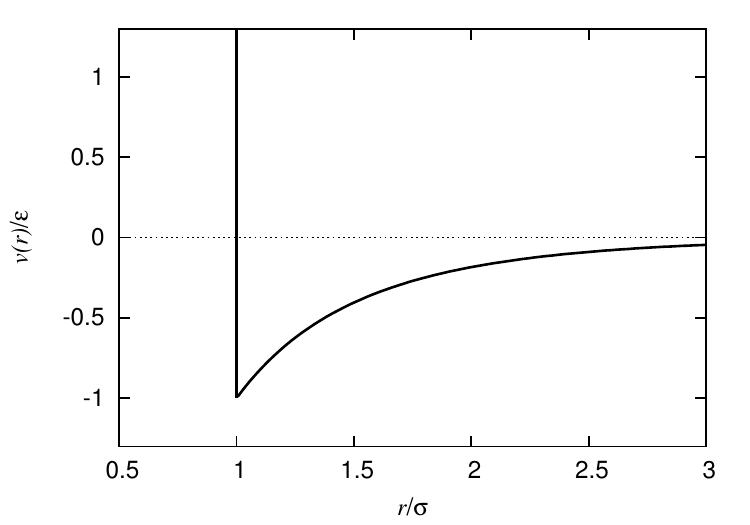}
\caption{\label{fig:pair_pot}{The Yukawa pair potential \eqref{eq:Yuk_pot}, with $\lambda=\sigma$, which is the interaction potential between the fluid particles in our system, plotted as a function of $r$, the distance between the centres of the particles. The parameter $\epsilon$ determines the strength of the attraction for $r>\sigma$, where $\sigma$ is the diameter of the (hard) cores of the particles.}}
\end{figure}

The model fluid that we consider consists of a system of particles interacting via a pair potential that can be split as follows:
\begin{equation}
v(r) = v_0(r)+v_1(r),
\end{equation}
where $r$ is the distance between the centres of the pairs of particles and $v_0(r)$, the repulsive-core part of the potential, is treated via the hard-sphere potential
\begin{equation}
 v_0(r)=\begin{cases}
    \infty & \text{if $ 0<r\leq \sigma $},\\
   0 & \text{if $ \sigma < r$},
  \end{cases}
\end{equation}
where $\sigma$ is the diameter of the cores of the particles. We model the attractive part of the potential $v_1(r)$ via the following Yukawa potential
\begin{equation}
 {v_1}(r)=\begin{cases}
   \enspace -\epsilon & \text{if $ 0<r\leq \sigma $},\\
  \enspace  \frac{-\epsilon e^{-(r-\sigma)/\lambda}}{r/\sigma} & \text{if $ \sigma < r$},
  \end{cases}\label{eq:Yuk_pot}
\end{equation}
where the range of the potential is defined by the length parameter $\lambda$ and the strength of the attraction is determined by the interaction energy parameter $\epsilon$. {A plot of the pair potential \eqref{eq:Yuk_pot} is displayed in Fig.~\ref{fig:pair_pot}, for $\lambda=\sigma$, the value used throughout this paper.} We use this Yukawa model potential because it is a widely studied model fluid in DFT, see e.g.\ Refs.~\cite{sullivan1979van, evans1986capillary, tarazona1987phase, louis2002effective, archer2013relationship} for a few examples from over the years, providing a good model for simple liquids \cite{hansen2013theory}.

\subsection{DFT implemented}

To treat this model fluid using DFT, we must develop an approximation for the excess Helmholtz free energy functional $F_{ex}$ in Eq.~\eqref{Fe}. We make a standard approximation, and treat the contribution to the free energy from the hard-sphere repulsions via fundamental measure theory (FMT) DFT and the attractive part via a van der Waals mean field like contribution \cite{evans1979nature, hansen2013theory, evans92, wu2006density, roth2010, PhysRevLett.63.980}, that is nonetheless fairly accurate \cite{archer2017standard}. Thus, the approximation we make is
\begin{equation}
F_{ex}[\rho(\mathbf r)] = F_{hs} + \frac{1}{2}\iint \rho(\mathbf {r}_1)\rho( \mathbf {r}_2) v_1 (\left|\mathbf {r}_1- \mathbf {r}_2 \right|)\mathrm d\mathbf {r}_1 \mathrm d\mathbf {r}_2,
\label{eq:F_ex}
\end{equation}
where $F_{hs}$ is the hard-sphere contribution to the free energy, that we treat using Rosenfeld's original version of FMT \cite{PhysRevLett.63.980}. There are more modern FMTs that are more accurate when the fluid density is high and approaching freezing \cite{hansen2013theory, roth2010, PhysRevLett.63.980}, but for the present study, Rosenfeld is sufficiently accurate.

\subsection{Bulk fluid phase diagram}

For bulk liquid-vapour coexistence to occur, the temperature $T$, pressure $p$ and chemical potential $\mu$ must be equal in the two coexisting phases. Substituting into Eqs.~\eqref{Fe} and \eqref{eq:F_ex} that the fluid density is a constant $\rho(\mathbf {r})=\rho=N/V$, where $N$ is the average number of particles in the system, and $V$ is the volume, then we obtain the Helmholtz free energy of the uniform fluid. The pressure is then obtained from this expression as the derivative $p=(\partial F/\partial V)_{N,T}$ and the chemical potential as $\mu=-(\partial F/\partial N)_{V,T}$. From these two relations, we can then write down a set of simultaneous equations for the coexisting vapour and liquid densities, $\rho_v$ and $\rho_l$, respectively, which are then solved for numerically over a range of temperatures to obtain the bulk fluid binodal \cite{hansen2013theory}.

 \begin{figure}[t]
 \includegraphics[width=1.\columnwidth]{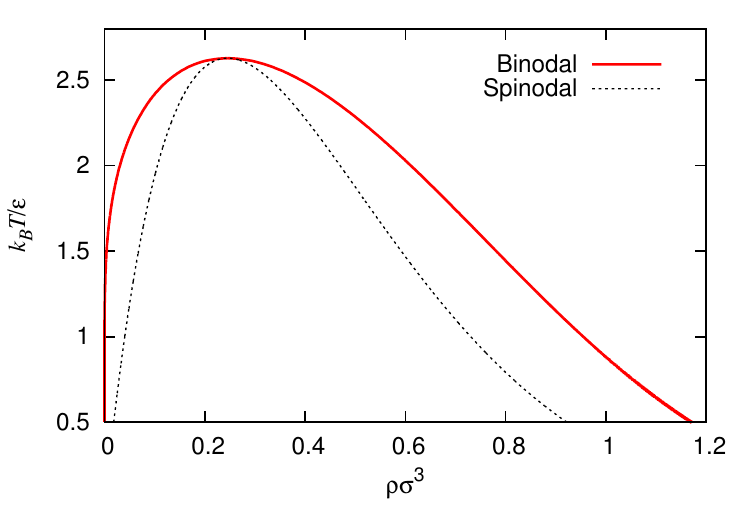} 
\caption{\label{phasediagram} Bulk fluid phase diagram in the temperature versus density plane, for the system with $\lambda=\sigma$. The solid line corresponds to the binodal curve and the dashed line corresponds to the spinodal curve.}
\end{figure}

In Fig.~\ref{phasediagram} we display the resulting bulk fluid phase diagram, showing the binodal curve giving the two distinct densities of the vapour and liquid phases at bulk coexistence. As the temperature $T$ is increased, the density difference between the two coexisting phases decreases and finally becomes zero at the critical temperature $T_c$. The fluid in the area of the phase diagram outside the binodal curve corresponds to the single phase region, where there is no phase separation. Inside is the two phase region, where vapour-liquid coexistence occurs. We also display the spinodal, which is given by the condition $
\partial^2 f/\partial \rho^2 = 0$, where $f=F/V$ is the free energy per unit volume. Inside the spinodal curve spontaneous phase separation occurs, whilst between the spinodal and the binodal, phase separation is a nucleated process, with a free energy barrier that must be surmounted by thermal fluctuations.

In the present work, we perform calculations at $k_BT/\epsilon=2$, which is sufficiently far from the critical point to see well separated bulk densities of $\rho_l \sigma^3 = 0.61$ and $\rho_v \sigma^3 =0.03$, { where at coexistence $\mu=\mu_{coex}$ the pressure $\beta\sigma^3p=0.026$}.

\subsection{External potential due to the wall}\label{subsec:ext_pots}

We assume that the planar solid substrate exerts an external potential on the fluid that varies in only one Cartesian direction, along the $z$-axis, which is perpendicular to the plane of the substrate. Having chosen to model the fluid particle-particle interactions via the Yukawa pair potential in Eq.~\eqref{eq:Yuk_pot}, an obvious choice for the potential between the particles and the wall is also a Yukawa:
\begin{equation}\label{Yukawall}
 {V_{ext}}^{(Y)}(z)=\begin{cases}
     \enspace \infty  & \text{if $z<\frac{\sigma}{2}$}
    \\
   \enspace  \frac{-\epsilon_{w}^{(Y)}e^{-z/\lambda_w^{(Y)}}}{z/\sigma}& \text{if $ z\geq\frac{\sigma}{2}$},
     \end{cases}
\end{equation}
where the parameters $\epsilon_{w}^{(Y)}$ and $\lambda_w^{(Y)}$ determine the strength of the attraction to the wall and the range, respectively.

We also consider the behaviour of the fluid in the presence of a wall with a {$z^{-3}$} power-law form for the decay of the attractive part of the potential. {Such a potential can be viewed as originating from the $r^{-6}$} decay form of the potential due to dispersion interactions that is found in e.g.\ the Lennard-Jones (LJ) model pair potential \cite{hansen2013theory}. {If one assumes a semi-infinite wall of uniform density and then integrates over the total attractive contribution due to the wall, treating all the elements as interacting with a given fluid particle with a potential decaying $\propto r^{-6}$, then} the resulting form is {(see e.g.\ Ref.~\cite{chacko2017solvent})}
\begin{equation}\label{LJwall}
 {V_{ext}}^{(LJ)}(z)=\begin{cases}
     \enspace \infty  & \text{if $z<\frac{\sigma}{2}$},
    \\
   \enspace  \frac{-\epsilon_{w}^{(LJ)}}{\left(z/\sigma\right)^3} & \text{if $z\geq\frac{\sigma}{2}$},
  \end{cases}
\end{equation}
where the parameter $\epsilon_{w}^{(LJ)}$ defines the strength of the attraction in this potential. Another wall potential that we consider is one with a short-ranged attraction, decaying with a Gaussian form \cite{archer2002wetting}
\begin{equation}
{V_{ext}}^{(G)}(z)=\begin{cases}
     \enspace \infty  & \text{if $z<\frac{\sigma}{2}$}
    \\
   \enspace  -\epsilon_{w}^{(G)}e^{-(z/\lambda_w^{(G)})^2}& \text{if $z\geq\frac{\sigma}{2}$},
  \end{cases}
\end{equation}
where the parameters $\epsilon_{w}^{(G)}$ and $\lambda_w^{(G)}$ define the strength and range of this potential. Finally, we also consider a wall potential that has exponential decay
\begin{equation}\label{expform}
{V_{ext}}^{(E)}(z)=\begin{cases}
     \enspace \infty  & \text{if $z<\frac{\sigma}{2}$}
    \\
   \enspace -\epsilon_{w}^{(E)}e^{-z/\lambda_w^{(E)}}& \text{if $z\geq\frac{\sigma}{2}$},
  \end{cases}
\end{equation}
with parameters $\epsilon_{w}^{(E)}$ and $\lambda_w^{(E)}$ determining the strength and range of the potential. The reason that we consider all these different potentials is that the form of the decay as $z\to\infty$ influences the form of the decay of $g(h)$ for $h\to\infty$ \cite{dietrich1988inphase, archer2002wetting}, as we also show below.

\begin{figure}[t]
 \includegraphics[width=1.\columnwidth]{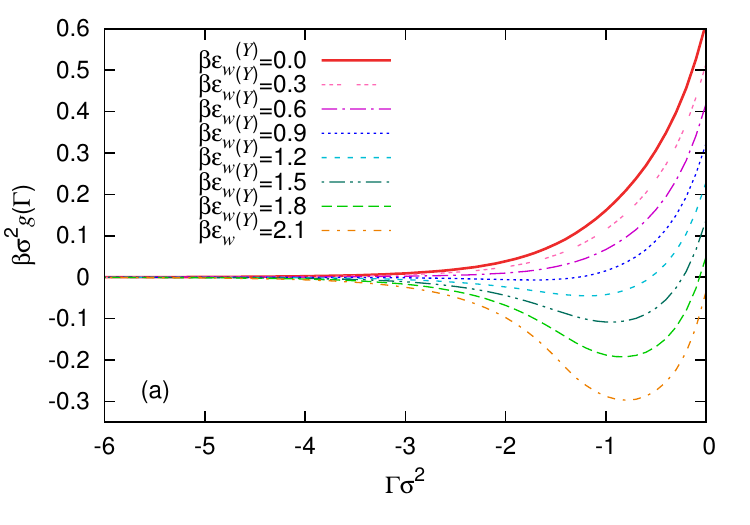}

 \includegraphics[width=1.\columnwidth]{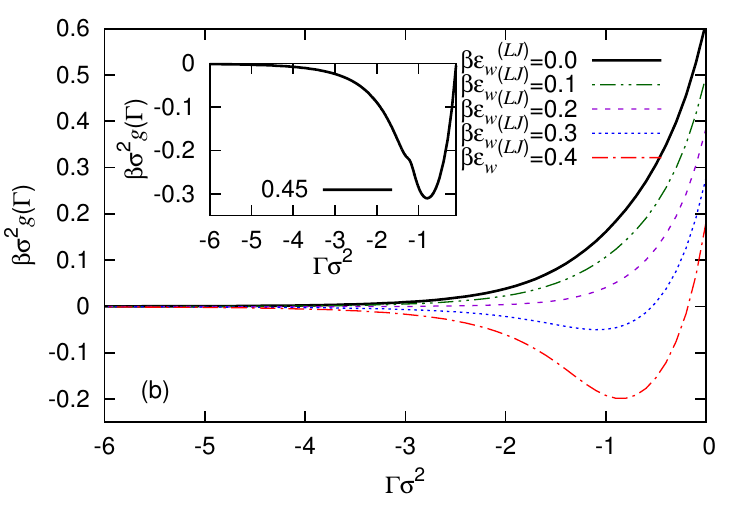}
\caption{\label{yukawa_vary_ew}A sequence of binding potentials $g(\Gamma)$, for varying wall attraction strength. The fluid pair interactions have $\beta\epsilon=0.5$ and $\lambda/\sigma=1$. In (a) we display results for the Yukawa wall potential (\ref{Yukawall}), for varying $\beta\epsilon_{w}^{(Y)}$ as given in the key, whilst in (b) are results for the LJ-like wall (\ref{LJwall}) with varying $\beta\epsilon_{w}^{(LJ)}$. The inset shows the binding potential for the strongly attractive wall with $\beta\epsilon_{w}^{(LJ)}=0.45$. In all except this last case the binding potentials are smooth and featureless, but in this case some small amplitude oscillations can be seen in $g(\Gamma)$. }
\end{figure}

{All our calculations of density profiles are performed on a regular grid with $2^{13}$ points and a grid spacing $dz=0.02\sigma$, so that the total domain length is $164\sigma$. This has the wall at one end of the system and a section at the other end with $\rho(z)=\rho_l$ (i.e.\ the bulk density boundary condition), followed by a section where $\rho(z)=0$, to provide padding for the fast Fourier transforms used to evaluate the convolution integrals. For more details on how to calculate density profiles using DFT see Ref.~\cite{roth2010}.}

\section{Results for the binding potential}\label{sec:results}

We calculate the binding potentials $g(\Gamma)$ for a range of different values of the adsorption $\Gamma$ using the procedure described above in Sec.~\ref{sec:DFT_approach}, for the various different wall potentials given in the previous section and for varying values of the attraction strength parameter. In Fig.~\ref{yukawa_vary_ew}(a) are results for the Yukawa wall potential (\ref{Yukawall}) and in Fig.~\ref{yukawa_vary_ew}(b) are results for the LJ-like wall potential (\ref{LJwall}). We see that in both cases, when the solid substrate is very weakly attractive, the global minimum of $g(\Gamma)$ is at $\Gamma \to -\infty$, corresponding to drying of the fluid from the wall being the equilibrium state of the system. For the more attractive substrates, the global minimum of the binding potentials is at a small negative value of the adsorption, which corresponds to the partial-drying situation. Our results are consistent with previous DFT predictions that the drying transition for these types of systems is a continuous (critical) transition -- see Ref.~\cite{evans2017drying} and references therein for an excellent recent discussion of this. It is interesting to note that this minimum in $g(\Gamma)$ is fairly broad and the binding potentials are rather smooth and featureless, despite the density profiles which go into calculating these having significant structure near the wall -- see Fig.~\ref{density_09}. The width of the minimum in $g(\Gamma)$ is certainly broader than the typical minima obtained in Ref.~\cite{hughes2017influence} for the case of liquid films adsorbed at a wall with the bulk phase being the vapour. We believe this is due to the fact that when there is the tendency towards drying at a solvophobic interface, there can be significant interfacial fluctuations \cite{jamadagni2011hydrophobicity, kanduc2016water, evans2015local, evans2015quantifying, evans2017drying, chacko2017solvent} and so in these cases any minima in $g(\Gamma)$ are fairly broad.

In the inset to Fig.~\ref{yukawa_vary_ew}(b) we show the binding potential for a more strongly attracting wall, with $\beta\epsilon_{w}^{(LJ)}=0.45$. In this case, the liquid is more strongly attracted to the wall and so we see more layered packing effects at the wall in the corresponding density profiles (not displayed). In cases like this, convergence of the numerics become more difficult, because the system does not want the vapour phase at the wall, since the liquid is energetically much more favourable. We also see in this situation the appearance of some small amplitude oscillations in the binding potential, stemming from particle layering at the wall.

\begin{figure}[t]
 \includegraphics[width=1.\columnwidth]{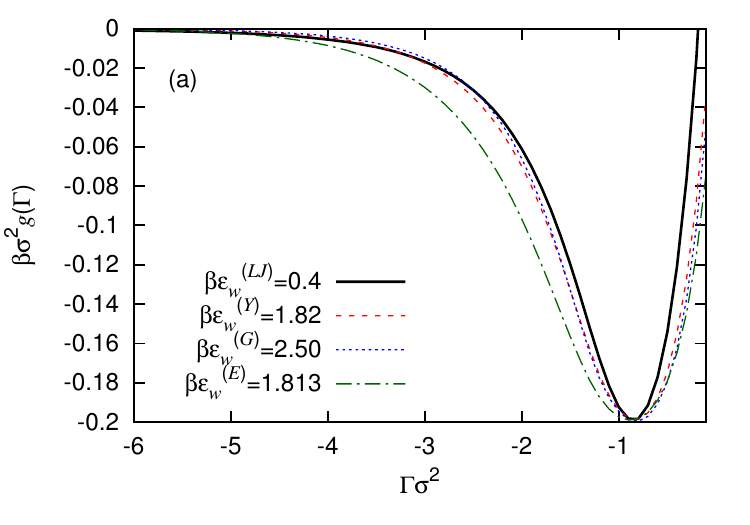}
 
 \includegraphics[width=1.\columnwidth]{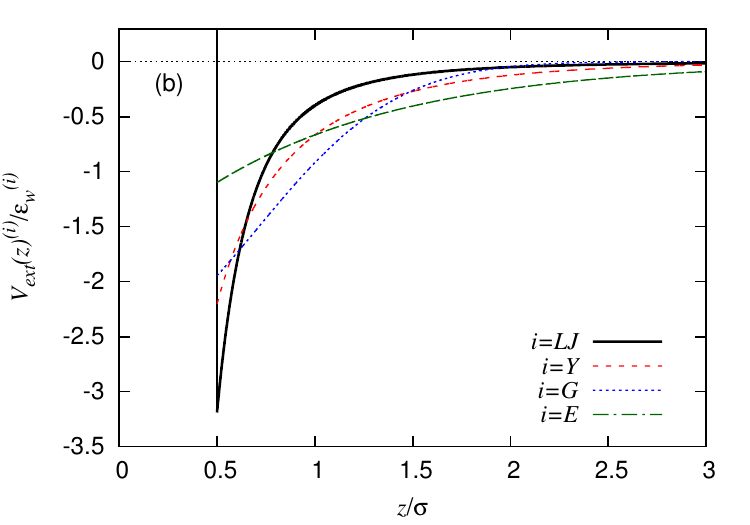}
\caption{\label{comparelj}{In panel (a) we show a} comparison of the binding potentials corresponding to {the} four different external potentials defined in Sec.~\ref{subsec:ext_pots}. The bulk fluid is the same in all cases, with $\beta\epsilon=0.5$ and $\lambda/\sigma=1$. The parameters are chosen as given in the key and with $\lambda_{w}^{(Y)}=\lambda_{w}^{(G)}=\lambda_{w}^{(E)}=\lambda$ (all the same), so that they all have the same minimal value of $g(\Gamma_0)$ and therefore also the same macroscopic contact angle. {In panel (b) we display plots of the corresponding four different wall potentials.}}
\end{figure}

\begin{figure}[t]
 \includegraphics[width=1.\columnwidth]{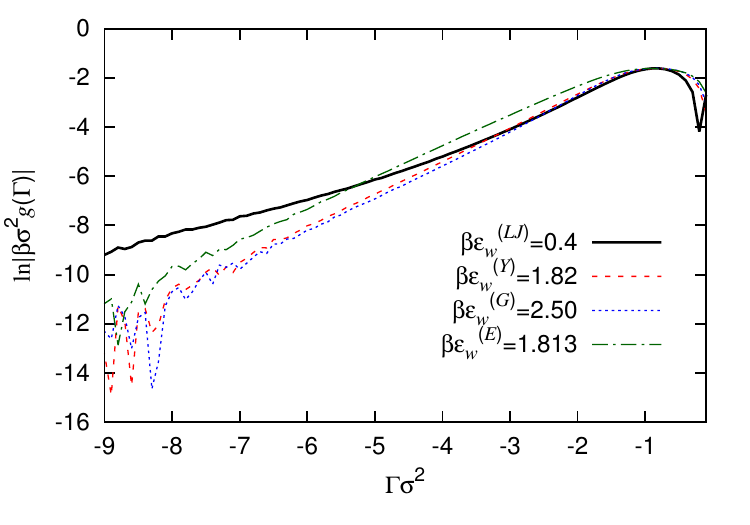}
\caption{\label{lg_comp} The same binding potentials as displayed in figure \ref{comparelj}{(a)}, except here we instead plot $\ln |g(\Gamma)|$ versus $\Gamma$.}
\end{figure}

In Fig.~\ref{comparelj}{(a)} we compare four binding potentials corresponding to the four different external potentials defined in Sec.~\ref{subsec:ext_pots}, with the wall potential attraction strength parameters chosen so that they all have the same minimal value of $g(\Gamma_0)$. Since the vapour-liquid interfacial tension $\beta\sigma^2 \gamma_{lv}=0.603$ is the same in all cases, this means that these all correspond to the same macroscopic contact angle, {because they all have the same minimum value of $g(\Gamma_0)$} -- see Eq.~\eqref{youngs}. It is interesting to note that the width of the potential minimum in $g(\Gamma)$ is not the same for each of these different wall potentials. This means the precise form of the external potential due to the wall is important for controlling the amplitude of interfacial fluctuations near the wall. We also see that the form of the external potential controls significantly the way $g(\Gamma)$ decays as $\Gamma\to-\infty$. This can be seen even more clearly in Fig.~\ref{lg_comp} where we instead plot $\ln |g(\Gamma)|$ versus $\Gamma$, which allows to observe more clearly the form of the asymptotic decay. The form of the asymptotic decay of binding potentials is discussed extensively in Refs.~\cite{dietrich1988inphase, schick1990liquids, henderson2005statistical}, and these results largely carry over to the case of drying at interfaces -- see Ref.~\cite{evans2017drying}. As one should expect, the slowest decay is for the LJ-like wall potential \eqref{LJwall}, since this has a power-law decay for $z\to\infty$. For the other three wall potentials the binding potential decays exponentially, so that when we plot $\ln|g(\Gamma)|$, we see in Fig.~\ref{lg_comp} a straight line. We see that the gradient is roughly the same for all three. This is because at this particular state point the correlation length in the vapour phase $\xi_v\approx \sigma=\lambda$, i.e.\ is very similar in value to the decay length of the wall potentials \eqref{Yukawall} and \eqref{expform}. For short-ranged wall-fluid and fluid-fluid potentials one should expect the binding potential to decay for $h\to\infty$ as \cite{dietrich1988inphase, schick1990liquids, archer2002wetting, evans2017drying}
\begin{equation}\label{eq:gg}
g(h)= a\exp(-h/\xi_v)+\cdots
\end{equation}
where $a$ is a constant and ``$\cdots$'' denotes faster decaying terms. So in this case, when one plots $\ln|g(\Gamma)|$, for large $\Gamma$ one sees a straight line with gradient equal to $-1/[\xi_v(\rho_v-\rho_l)]$. On the other hand, if there is an exponentially decaying wall potential \eqref{expform}, then one instead has \cite{archer2002wetting}
\begin{equation}\label{eq:ggg}
g(h)= a\exp(-h/\xi_v)+b\exp(-h/\lambda_w^{(E)})+\cdots,
\end{equation}
where $b$ is a constant, so whichever is bigger out of $\xi_v$ and $\lambda_w^{(E)}$ determines the ultimate decay of $g(h)$ for $h\to\infty$. When the wall potential has a Yukawa decay like in Eq.~\eqref{Yukawall}, then this can also determine the decay of $g(h)$, somewhat like in Eq.~\eqref{eq:ggg}, except with a renormalised decay length \cite{archer2002wetting}. {Note that for larger negative values of the adsorption the binding potential $g(\Gamma)$ becomes small and so on the logarithmic scale in Fig.~\ref{lg_comp} one sees the numerical round-off errors, appearing as random fluctuations with increasing amplitude as $\Gamma\to-\infty$.}

{It is also interesting to note in Fig.~\ref{comparelj}(a) that all of the binding potentials have a finite value for $g(\Gamma\to0)$, but the values of $g(0)$ for the different wall potentials are all very different and in particular the result corresponding to the LJ wall is much higher. We believe the origin of this difference is the fact that the LJ wall potential \eqref{LJwall} has a deeper (but more narrow) potential minimum for $z\to\sigma/2^+$ than the other wall potentials, as can be seen in Fig.~\ref{comparelj}(b). This is also supported by the fact that the values of $g(0)$ are ordered in magnitude in the same order as the values of the wall potentials at contact, $V_{ext}^{(i)}(z\to\sigma/2^+)$. That the value of $g(0)$ must be finite was discussed in the context of liquid droplets at surfaces in Refs.~\cite{hughes2015liquid, hughes2017influence}. Indeed, $g(\Gamma)$ remains finite even for small positive values of $\Gamma$, which corresponds to a negative excess of vapour being adsorbed at the wall. However, the fact that $g(0)$ remains finite should not significantly affect the behaviour at the contact line, since the value at the minimum $g(\Gamma_0)$ is far more important than the value $g(0)$ in determining contact line properties.}

\begin{figure}
 \includegraphics[width=1.\columnwidth]{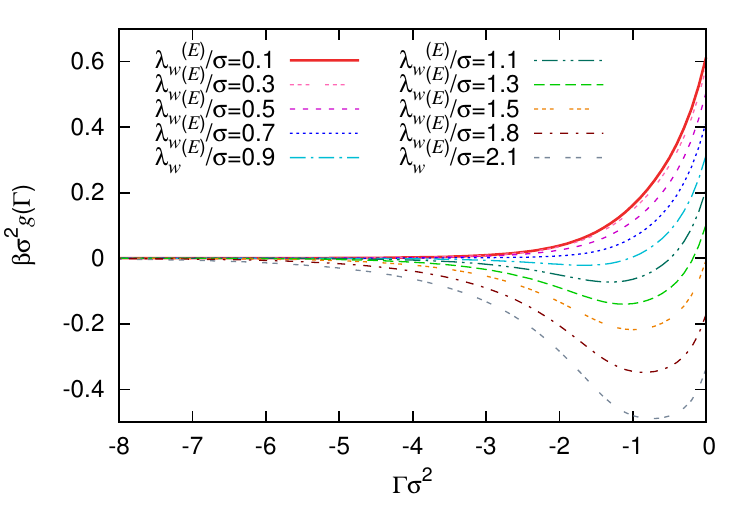}
\caption{\label{comparelambda_exp}A series of binding potentials for the exponential wall potential \eqref{expform} with varying $\lambda_w^{(E)}$ and fixed $\beta\epsilon_w^{(E)}=1$. The fluid pair interactions have $\beta\epsilon=0.5$ and $\lambda/\sigma=1$.}
\end{figure}

\begin{figure}
 \includegraphics[width=1.\columnwidth]{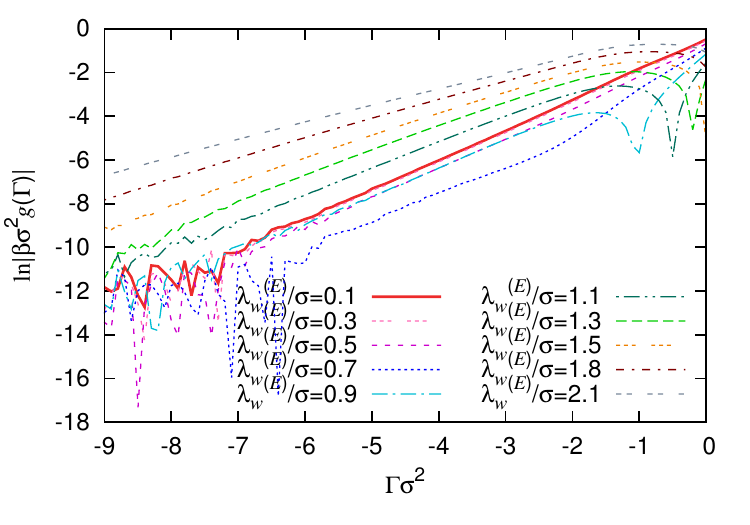}
\caption{\label{log_exp}\label{lg_comp2} The same binding potentials as displayed in Fig.~\ref{comparelambda_exp} for varying $\lambda_w^{(E)}$, but here we instead plot $\ln |g(\Gamma)|$ versus $\Gamma$.}
\end{figure}

In Fig.~\ref{comparelambda_exp} we display a set of binding potentials for the exponential wall potential \eqref{expform}, calculated for varying wall potential decay length $\lambda_w^{(E)}$. Increasing the range for fixed $\epsilon_w^{(E)}$ increases the overall integrated strength of the wall potential and so, of course, makes the liquid more favourable at the wall and the vapour less favourable. This is manifest in the increasingly deep minimum in $g(\Gamma)$, as $\lambda_w^{(E)}$ is increased. In Fig.~\ref{log_exp} we plot $\ln|g(\Gamma)|$, which allows one to see the crossover from the first term on the right hand side of Eq.~\eqref{eq:ggg} dominating the decay of $g(\Gamma)$, to the second term dominating, for larger $\lambda_w^{(E)}$.

In the following section we take the binding potentials that we have calculated using DFT and input them into the IH \eqref{IH} in order to determine vapour nanobubble height profiles. To do this we fit the binding potential to obtain an analytic form which can then be input easily. The form we use is (c.f.\ Eq.~\eqref{eq:ggg} and also Refs.~\cite{hughes2015liquid, hughes2017influence}):
\begin{equation}
\label{eq:fit_g}
g(\Gamma) = a_1e^{\frac{\Gamma}{l_0}}+a_2e^{\frac{2\Gamma}{l_0}}+a_3e^{\frac{3\Gamma}{l_0}}+\cdots
\end{equation}
where $l_0$, $a_1$, $a_2$, $a_3$, etc, are parameters to be fitted. {The values obtained for these parameters for all of the binding potentials displayed in this paper are given in Table~\ref{table:nonlin} in the Appendix.} Recall that $\Gamma$ is normally a negative quantity in Eq.~\eqref{eq:fit_g}.

\section{Vapour nanobubble profiles}\label{sec:bubble_profiles}

\begin{figure}
 \includegraphics[width=1.\columnwidth]{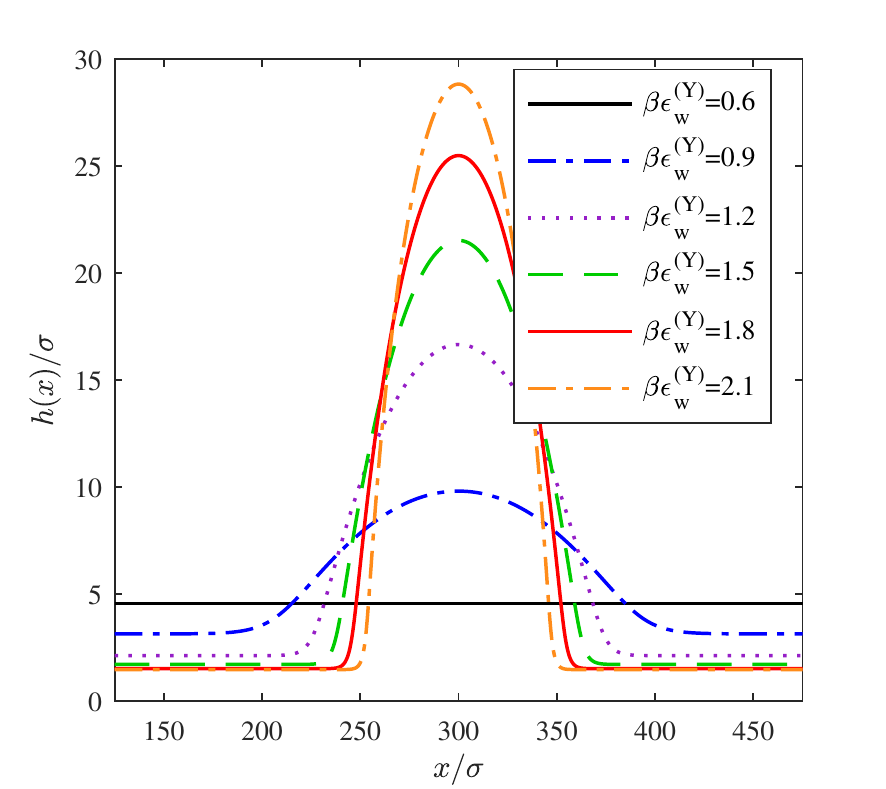}
\caption{\label{comparebubbleprofile} A series of equilibrium vapour nanobubble height profiles $h(x)=\Gamma(x)/(\rho_v-\rho_l)$, calculated by minimising Eq.~\eqref{IH} together with the binding potentials for the fluid with $\beta\epsilon=0.5$ and $\lambda/\sigma=1$ at the Yukawa wall \eqref{Yukawall}, with fixed $\lambda_{w}^{(Y)}/\sigma=1$ and various values of the wall attraction parameter $\epsilon_{w}^{(Y)}$, as given in the key. The total area under all of the curves is $2727\sigma^2$ and the length of the domain $L=600\sigma$.}
\end{figure}

In Fig.~\ref{comparebubbleprofile}, we display a sequence of equilibrium vapour nanobubble height profiles $h(x)=\Gamma(x)/(\rho_v-\rho_l)$, calculated by minimising Eq.~\eqref{IH} together with binding potentials calculated using DFT. We do this for the fluid with interaction parameters $\beta\epsilon=0.5$ and $\lambda/\sigma=1$ at a series of walls with the Yukawa potential \eqref{Yukawall} with fixed $\lambda_{w}^{(Y)}/\sigma=1$ and various values of the wall attraction parameter $\epsilon_{w}^{(Y)}$. In Eq.~\eqref{IH} we set the liquid-vapour interfacial tension $\beta\sigma^2\gamma_{lv}=0.603$, the value we obtain from the DFT. We also assume for simplicity that the system is uniform in the $y$-direction, so strictly speaking the profiles that we calculate are actually for ridge-shaped nanobubbles. However, we do not expect results from calculating radially symmetric height profiles (varying in both the $x$- and $y$-directions) to have cross-section height profiles qualitatively different from the ones we calculate here. We apply periodic boundary conditions $h(x=0)=h(x=L)$, where $L$ is the length of the domain. The height profiles in Fig.~\ref{comparebubbleprofile} all have the same area under the curve (i.e.\ the same total adsorption).

\begin{figure}[t]
   \includegraphics[width=1.\columnwidth]{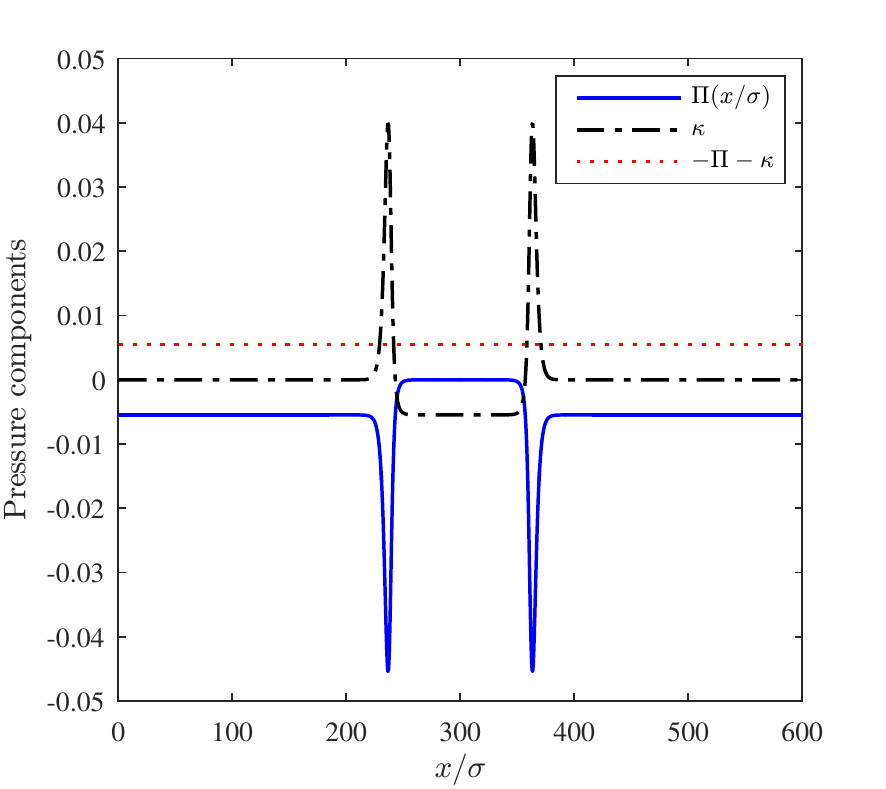} 
   \caption{The components of the excess pressure, $\Pi$ and $\kappa$, given by Eqs.~\eqref{eq:pressure_comps} and \eqref{eq:kappa}, for a nanobubble with volume $2727\sigma^2$ and wall attraction strength $\beta\epsilon_w^{(Y)}=1.5$. The corresponding height profile is displayed in Fig.~\ref{comparebubbleprofile}.}
   \label{fig:press_components}
\end{figure}

We numerically minimise the free energy \eqref{IH} by solving the corresponding thin-film equation with disjoining pressure $\Pi(h)=-\partial g/\partial h$ and converging to equilibrium, based on the approach of Ref.~\cite{yin2017films}. This uses the method of lines, with finite difference approximations for the spatial derivatives and the {\it ode15s} Matlab variable-step, variable-order solver \cite{matlabodesuite}. The initial guess to equilibrate from has a Gaussian shaped ``bump'' in it that breaks the symmetry and determines the final location of the nanobubble on the surface. In Fig.~\ref{comparebubbleprofile} we see that the vapour nanobubbles become more spread out over the surface as the attraction due to the wall is decreased. Then, for $\beta\epsilon_{w}^{(Y)}=0.6$, there is a uniform thickness film of vapour on the substrate. This corresponds to the drying transition and it occurs at the value of $\epsilon_{w}^{(Y)}$ that one must expect from inspecting the binding potential curves in Fig.~\ref{yukawa_vary_ew}(a), i.e.\ where the minimum in $g(h)$ at a finite value of $h$ disappears, {which occurs by the minimum value diverging $h\to\infty$, since this drying transition is continuous (critical).} For the profiles containing a nanobubble, the height of the vapour ``precursor'' film corresponds roughly to the value at the minimum in the binding potentials for the different values of $\beta\epsilon_{w}^{(Y)}$. However, in a finite size domain, the height is shifted slightly from the minimum value due to the Laplace pressure in the nanobubbles combined with the effects of mass conservation in our periodic domain. {The excess pressure due to the presence of the nanobubble has two components,
\begin{equation}\label{eq:pressure_comps}
\frac{\delta F_{\textrm{IH}}}{\delta h}=-\Pi(h(x))-\kappa(h(x)),
\end{equation}
where $F_{\textrm{IH}}$ is given in Eq.~\eqref{IH}, $\Pi$ is the disjoining pressure and the curvature contribution is
\begin{equation}\label{eq:kappa}
\kappa=\gamma_{lv}\nabla\cdot\left(\frac{\nabla h}{\sqrt{1+(\nabla h)^2}}\right).
\end{equation}
In Fig.~\ref{fig:press_components} we display the values of these two contributions to the excess pressure as a function of position through a nanobubble, for the case where $\beta\epsilon_w^{(Y)}=1.5$. The corresponding nanobubble height profile is displayed in Fig.~\ref{comparebubbleprofile}. We see in Fig.~\ref{fig:press_components} that these two pressure components vary significantly with $x$, in particular in the contact line region. Of course, the sum of these is a constant as this is the condition for equilibrium.}

\begin{figure}[t]
 \includegraphics[width=1.\columnwidth]{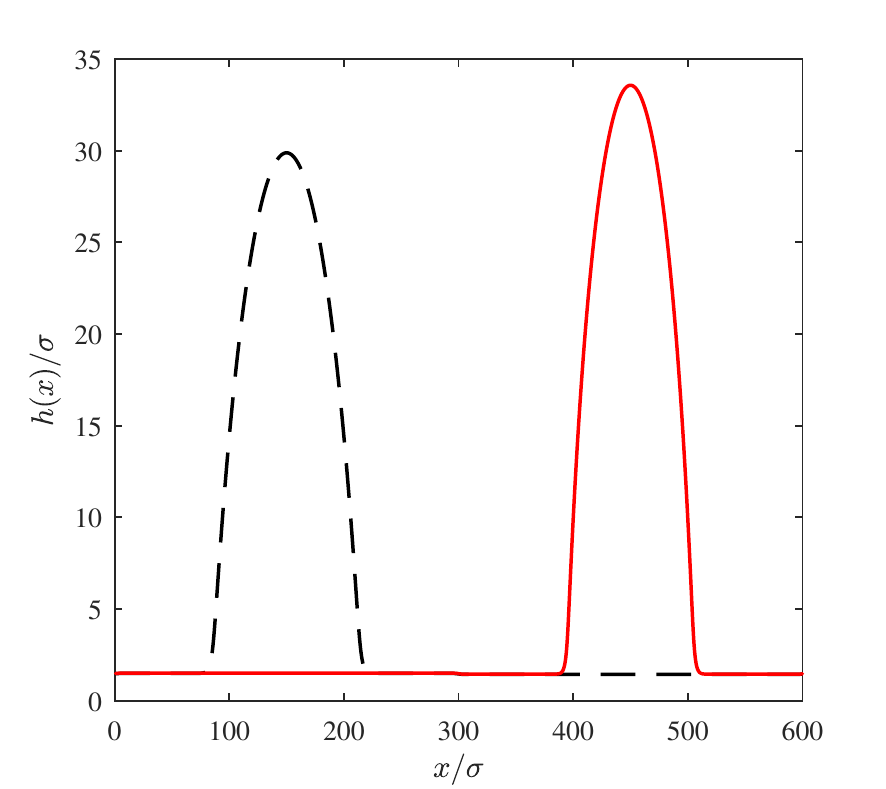}
\caption{\label{leftright} A comparison of two equilibrium vapour nanobubble profiles on a heterogeneous surface with position dependent binding potential \eqref{eq:x_var_g}. The external potential due to the wall has attraction strength $\beta\epsilon_w^{(Y)}=2.1$ on the right half of the system and $\beta\epsilon_w^{(Y)}=1.8$ on the left half. The total volume of vapour in the system is the same in both cases.}
\end{figure}

As an example of the type of multiscale interfacial phenomenon that our coarse grained model can be used to describe, we compute vapour nanobubble height profiles on a patterned heterogeneous surface. This consists of a surface divided into two regions with a different wettability on each of the two halves of the surface. We calculate the free energies for nanobubbles on each half, and from this we are able to determine the relative probabilities for finding vapour nanobubbles on each type of surface. We define our position dependent binding potential as
\begin{equation}\label{eq:x_var_g}
 g(x,h)=g_l(h)(1-f(x))+g_r(h)f(x),
 \end{equation}
 where the smooth switching function
  \begin{align}
 f(x) =& \frac{1}{2}\left[\tanh\left(\frac{x-L/2}{\mathcal{W}}\right)
 -\tanh\left(\frac{x-L}{\mathcal{W}}\right)\right]\nonumber\\
&+ \frac{1}{2}\left[\tanh\left(\frac{x+L/2}{\mathcal{W}}\right)
 -\tanh\left(\frac{x}{\mathcal{W}}\right)\right],
 \end{align}
where $\mathcal{W}=\sigma$ determines the width of the smooth transition zone between the two halves of the surface. This function also satisfies our periodic boundary conditions. $g_l(h)$ and $g_r(h)$ are the binding potentials on the left and right hand halves of the surface, respectively. These are calculated for the Yukawa wall with $\lambda_{w}^{(Y)}/\sigma=1$. On the right we have $\beta\epsilon_{w}^{(Y)}=2.1$, which represents a more solvophilic surface, whilst on the left we have a lower attraction parameter, $\beta\epsilon_{w}^{(Y)}=1.8$, which represents a more solvophobic surface.
 
\begin{figure}[t]
 \includegraphics[width=1.\columnwidth]{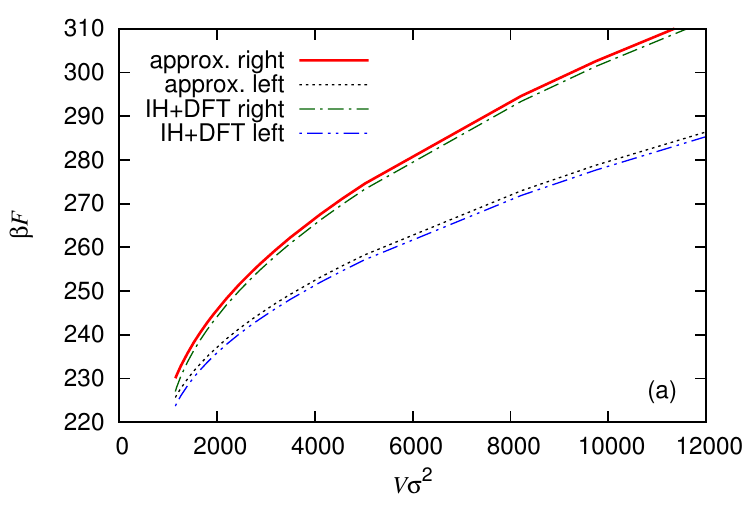}
 
  \includegraphics[width=1.\columnwidth]{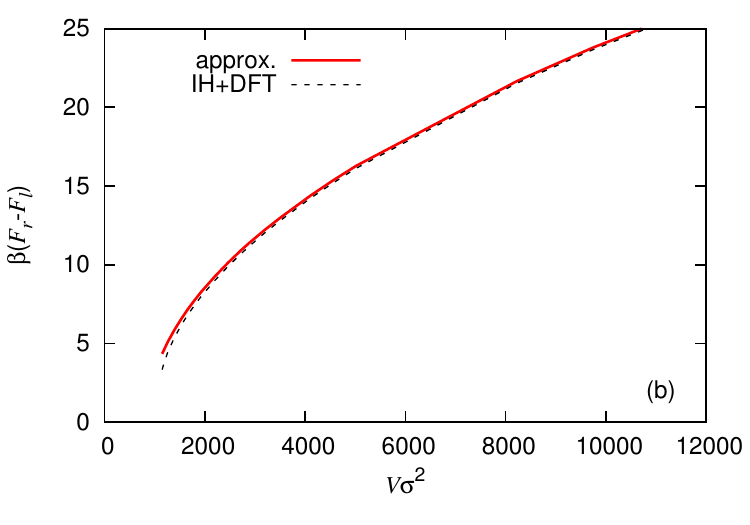}
\caption{\label{energy}{In panel (a) we display the} free energy $F$ as a function of the vapour nanobubble volume $V$, for a heterogeneous system with wall attraction $\beta\epsilon_{w}^{(Y)}=2.1$ on the right half of the surface and $\beta\epsilon_{w}^{(Y)}=1.8$ on the left. The labels ``right'' and ``left'' in the key denote on which side of the system the nanobubble is located -- c.f.\ Fig.~\ref{leftright}. We compare results calculated from Eq.~\eqref{IH} with the binding potentials obtained from DFT, which are labelled ``DFT+IH'' with results from a simple macroscopic approximation \eqref{eq:approx}, labelled ``approx.''. 
{In panel (b)} we plot the quantity {${\beta(F_r-F_l)}$ as a function of $V$. The exponential $e^{-\beta(F_r-F_l)}$} gives the ratio of the probabilities $P_r/P_l$ of finding the nanobubble on the two sides. Since the left half of the surface is more solvophobic, we have $P_l>P_r$.}
\end{figure}

In Fig.~\ref{leftright} we display the height profiles for two different nanobubbles having the same volume $V$ but each centred on the two different halves of the system. The total domain length is $L=600\sigma$. The initial condition used to calculate each of these has the Gaussian bump centred at either $x=L/4$ or $x=3L/4$, in order to locate the centres of the final equilibrium nanobubbles at these points. The left hand vapour nanobubble on the less attractive wall (smaller $\epsilon_{w}^{(Y)}$) has the lower free energy. The free energy of the whole system $F$ is calculated using Eq.~\eqref{IH} and in Fig.~\ref{energy}{(a)} we display results for $F$ calculated as a function of $V$. In this figure these results are compared with those from a simple macroscopic (capillarity) approximation, described below. Using this data, in {Fig.~\ref{energy}(b) we plot the quantity $\beta(F_r-F_l)$} as a function of $V$, where $F_l$ is the free energy for the nanobubble on the left and $F_r$ when it is on the right. Since the probability of a given state $i$ occurring $P_i\propto e^{-\beta F_i}$, we therefore have that the ratio of the probabilities for finding the nanobubble on the two different halves of the system $P_r/P_l=e^{-\beta(F_r-F_l)}$, {i.e.\ the exponential of minus the quantity displayed in Fig.~\ref{energy}(b) is the relative probability}. Since the left half of the surface is more solvophobic, we have $P_l>P_r$, and as the size of the nanobubbles increases, the probability of finding such a nanobubble on the more solvophobic half of the system becomes much more likely, with the relative probability, $P_l\gg P_r$. Note that the curves in Fig.~\ref{energy} end on the left at a finite value of the volume $V$. This is because when the volume of vapour in the system is less than the end point value, the system can lower the total system free energy by having a uniform film thickness everywhere, at a value shifted slightly from the value at the minimum of $g(h)$, rather than by having most of the system with $h$ at the minimum of $g(h)$ but also retaining a bubble which has a larger interfacial contribution from curvature.

The macroscopic (capillarity) approximation that we compare our results with consists of setting the height profile of the vapour nanobubble to be an analytic piecewise function of $x$. We assume that outside of the nanobubble the film height is uniform: in the left half of the system we set $h(x)=h_l$, where $h_l$ is the value at the minimum of the binding potential $g_l(h)$ and in the right half we set $h(x)=h_r$, where $h_r$ is the value at the minimum of $g_r(h)$. For the nanobubble itself, we assume the height profile is the {arc of a circle $h(x)=h_{circ}(x)=h_c+\sqrt{R^2-(x-x_c)^2}$, where $h_c$, $x_c$ and $R$ are constant coefficients to be determined that depend on the size and location of the nanobubble}. If we denote the locations of the two nanobubble contact lines to be $x=A$ and $x=A+w$, i.e.\ $w$ is the width of the nanobubble, then {$x_c=A+w/2$ and} we must have that at these two points, the height profile is continuous. So, when the nanobubble is on the left we have $h(A)=h(A+w)=h_l$ and when it is on the right, $h(A)=h(A+w)=h_r$. The second condition that we apply on the {circular arc part of the nanobubble profile is that the slope at both ends should be equal to the tangent of the contact angle, $h'(A)=-h'(A+w)=-\tan\theta$. With these conditions, it is straightforward to write the coefficients $R$ and $h_c$} as functions of $A$ and $w$. When the nanobubble is on the left hand side, the volume (area under the profile) is:
\begin{equation}
V = h_l\left(\frac{L}{2}-w\right)+h_r\frac{L}{2}+\int_{A}^{A+w} h_{circ}(x) \mathrm{d}x,
\end{equation}
with an analogous formula for when it is on the right. This gives us an expression for $V$ as a function of $w$. Or, equivalently, we can vary $w$ and still obtain a series of nanobubble profiles for various values of $V$.

Using this height profile we can also obtain an approximation for the free energy $F$. The surface tension contribution depends on the length of the interface. This is easy to get for the straight line pieces and for the {circular} nanobubble section it depends on the arc length
\begin{equation}
s=\int_{A}^{A+w}\sqrt{(1+h'(x)^2)} \mathrm{d}x,
\end{equation}
which is also straightforward to evaluate. We assume that there is only a contribution to $F$ from the binding potential when the height profile is at the value at the minimum of $g(h)$. Putting all this together we obtain the following estimate for the total free energy of the system when the vapour nanobubble is on the left hand side of the system
\begin{equation}
F_{\textrm{IH}}^{\textrm{approx.}}=g_l(h_l)\left(\frac{L}{2}-w\right)+g_r(h_r)\frac{L}{2}+\gamma_{lv}(s+L-w),
\label{eq:approx}
\end{equation}
and an analogous expression when the nanobubble is on the right. The results plotted in Figs.~\ref{energy} labelled ``approx.'' are obtained using Eq.~\eqref{eq:approx}. We see that there is fairly good agreement in Fig.~\ref{energy}{(a)} between Eq.~\eqref{eq:approx} and the results from the full minimisation of Eq.~\eqref{IH}; the difference is less than {1\%. However, as Fig.~\ref{energy}(b) illustrates}, even such small errors can make {more of a difference when calculating quantities like $(F_r-F_l)$ and so also the ratio $P_r/P_l=e^{-\beta(F_r-F_l)}$}, demonstrating the importance of getting details right for this sort of calculation. {This is particularly important for small nanobubbles. For example, when the nanobubble volume $V\sigma^2=1400$, we have $e^{-\beta(F_r-F_l)}=e^{-5.9}\approx0.0027$ via Eq.~\eqref{eq:approx}, but from the full minimisation of Eq.~\eqref{IH} we obtain $e^{-\beta(F_r-F_l)}=e^{-5.4}\approx0.0045$; i.e.\ there is a 60\% difference between the two results for the relative probabilities $P_r/P_l$. Another important detail for these types of calculations} is getting correctly the true overall shape of $g(h)$, since this makes a contribution to $F$, which is neglected in Eq.~\eqref{eq:approx}, coming from the contact line region of the nanobubble.

Another source of error in Eq.~\eqref{eq:approx} worth highlighting is that we have assumed that the heights of the film away from the nanobubble are the values at the exact minima of the binding potentials $g_l$ and $g_r$. Consequently, any additional vapour volume in the system is assumed to be in the nanobubble. In reality, as we see from results from minimising Eq.~\eqref{IH} and magnifying in the small $h$ region (not displayed), there is a balance between having the vapour in the small-$h$ flat layer and having it in the nanobubble. The Laplace pressure in the nanobubble makes it become a little smaller, transferring some of the vapour into the flat film and thereby raising the free energy contribution from these portions of the system. There are also further sources of error due to the assumption that the nanobubble has a {circular} shape, in particular in the region near the contact lines where it would be expected to smoothly transition to the film heights, and in the error approximating the profile's transition across the wettability gradient as a sharp step.

\begin{figure}[t] 
   \includegraphics[width=1.\columnwidth]{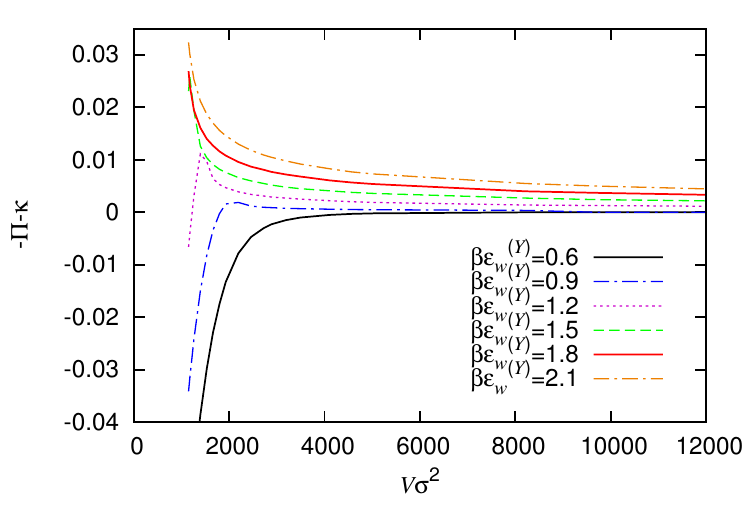} 
   \caption{The excess pressure $-\Pi-\kappa$, given by Eq.~\eqref{eq:pressure_comps}, for a range of different nanobubble volumes and for values of the wall attraction strength parameter $\beta\epsilon_w^{(Y)}$ as given in the key. See also Fig.~\ref{comparebubbleprofile}.}
   \label{fig:ex_press}
\end{figure}

{In Fig.~\ref{fig:ex_press} we display $(-\Pi-\kappa)$, the excess pressure due to the presence of the nanobubble, given by Eq.~\eqref{eq:pressure_comps}, for a range of different nanobubble volumes and for a range of different values of the wall attraction strength parameter $\beta\epsilon_w^{(Y)}$. Recall that the bulk fluid pressure is $\beta\sigma^3p=0.026$, so the figure shows that these excess pressures are comparable in magnitude. For values of $\beta\epsilon_w^{(Y)}$ smaller than that at the drying transition we see that for $V\to\infty$, $(-\Pi-\kappa)\to0$ from below, whilst for $\beta\epsilon_w^{(Y)}$ greater than that at the drying transition, then $(-\Pi-\kappa)\to0$ from above.}

\section{Concluding remarks}\label{sec:conc}

In this paper we have presented results for the binding potential $g(h)$ for films of vapour intruding between a bulk liquid and flat planar surfaces and used the calculated $g(h)$ to determine film height profiles for vapour nanobubbles on the surface. The binding potentials are calculated using a microscopic DFT, applying the fictitious external potential method developed by Hughes \emph{et al.}\ \cite{hughes2015liquid, hughes2017influence}, which is based on calculating a series of constrained fluid density profiles at the wall with varying thickness (adsorption). We see from our results in e.g.\ Fig.~\ref{comparelj}{(a)} that despite the resulting binding potentials being rather smooth and featureless, details such as the width of the minimum and the form of the decay in $g(h)$ do depend crucially on the details of the microscopic interactions. We also see from our estimates of the relative probabilities of finding a nanobubble on different parts of a heterogenous surface displayed in inset to Fig.~\ref{energy} that having a reliable approximation for $g(h)$ is necessary for the estimates to be accurate. It is clear that to correctly describe vapour nanobubbles one must have an accurate binding potential. Here, we have used a microscopic DFT based on FMT to determine $g(h)$, although one could instead use computer simulations \cite{macdowell2011computer, tretyakov2013parameter, benet2016premelting, jain2019using}. However, the DFT calculations are computationally much faster.

The overall coarse-graining procedure developed here, building on the work in Refs.~\cite{hughes2015liquid, hughes2017influence}, allows us to determine multi-scale properties of fluids at interfaces. The approach allows to go from the microscopic features of the molecular interactions and go up in length scales to describe mesoscopic aspects such as nanobubbles on surfaces. Our approach has here been applied to a simple model heterogenous surface, but it could also be applied in a straight-forward manner to more complex surfaces and structures, since, for example, the contributions to $g(h)$ from surface curvature are understood \cite{stewart2005critical, stewart2005wetting}.

In the work presented here we have assumed that it is just the vapour phase inside the nanobubbles. However, as mentioned in the introduction, perhaps the more experimentally relevant situation is when the nanobubbles also contain dissolved gas (i.e.\ air) molecules that have come out from solution in the bulk liquid. In Ref.~\cite{svetovoy2016effect} a theory for this situation is developed. The authors argue that one should set the binding potential in Eq.\ \eqref{IH} to be the potential $U(h)=w(h)-w(h_c)-\beta\mu_g h p_g(h)$, where $w(h)$ is the ``bare'' binding potential between the wall and the bulk liquid and the last term is the contribution from the gas in the nanobubble, that has chemical potential $\mu_g$ and pressure $p_g(h)$, which is assumed to be related to the disjoining pressure and given by the ideal-gas equation of state. Whilst this approach has the advantage of being relatively simple, one could also include the effects of dissolved gas in the present approach by treating the system as a binary mixture and then using a DFT for the mixture to determine the influence of different amounts of the gas at the interface on $g(h)$. Such a DFT approach would, of course, include the effects of the gas compressibility, which are believed to be important for such surface nanobubbles.

Finally, we should remark that some of the values of the wall attraction parameter $\epsilon_w^{(Y)}$ that we use are rather small, corresponding to very solvophobic surfaces. Considering simple molecular liquids at interfaces, such values are perhaps somewhat unrealistic, being weaker than one would typically expect to find. For example, for water on hydrophobic surfaces such wax or Teflon, one does not see contact angles significantly greater than 130$^\circ$ \cite{evans2017drying}. However, at (patterned) superhydrophobic surfaces much larger contact angles are possible, so studying the behaviour of the model right up to the drying transition is relevant to such systems. Also, the model fluid considered here is also a reasonably good model for certain colloidal suspensions (e.g.\ colloid-polymer mixtures \cite{dijkstra1999phase}) and for such systems even purely repulsive wall potentials are possible, when e.g.\ polymers are grafted onto the walls. The work here is highly relevant to such colloidal systems.

\section*{Acknowledgements}

We gratefully acknowledge Uwe Thiele for stimulating discussions and also for hosting AJA in M\"unster where some of this was written. DNS acknowledges support via EPSRC grant number EP/R006520/1.

{
\section*{Appendix}

{In Table \ref{table:nonlin} we give values of the coefficients in the binding potential $g(\Gamma)$ in Eq.~\eqref{eq:fit_g}, obtained by fitting to the results from DFT for a range of different values of the parameters in the wall potential, for the fluid with $\lambda=\sigma$ and $\beta\epsilon=0.5$.}

}
\begin{table*}[ht]
\caption{{The parameter values $a_1$, $a_2$, ..., $a_8$ and $l_0$ in the binding potential $g(\Gamma)$ in Eq.~\eqref{eq:fit_g}, obtained from fitting to the data calculated using the DFT, for the various different wall potentials given in Eqs.~\eqref{Yukawall}--\eqref{expform}. The attraction and range parameters $\epsilon_w^{(i)}$ and $\lambda_w^{(i)}$ in these potentials are also given below. The first column refers to the number of the figure above in which the binding potentials are displayed.}} 
\centering 
\begin{tabular}{c| c| c| c| c| c| c| c| c| c| c| c| cl} 
\hline\hline 
Figure & wall type & $\beta\epsilon_w^{(i)}$&$\lambda_w^{(i)}/\sigma$ &$a_1$&$a_2$&$a_3$&$a_4$&$a_5$&$a_6$&$a_7$&$a_8$&$l_0$ \\ [0.5ex] 
\hline 
3  & Y & 1.817 &1& -0.102902&-1.52976&-7.19867&45.6063&-82.5011&64.6922&-18.9215&0&1.13494\\ 
6(a) & Y & 0 &1& 0.436017&1.56668&-5.80142& 10.5037&-10.2276&5.2632&-1.12803&0&0.764648\\
6(a) & Y &0.3&1&0.188742&1.01604&-1.10085&-1.90374&6.14653&-5.55662&1.72492&0&0.834599\\
6(a) & Y &0.6&1&0.0636616&-0.0223561&3.88825&-12.2983&17.6613&-12.1129&3.23861&0&0.898494\\
6(a) & Y &0.9&1&-0.00913047&-1.07813&8.16947&-20.2538&25.6447&-16.2729&4.12271&0&0.914339\\
6(a) & Y &1.2&1&-0.103283&-2.23702&13.5188&-31.1874&37.6513&-23.154&5.74073&0&0.894094\\
6(a) & Y &1.5&1&-0.316933&-3.50244&22.3324&-53.6347&66.6376&-42.0627&10.6847&0&0.832845\\
6(a) & Y &1.8&1&-0.0998242&-1.4847&-6.998&44.2127&-79.8251&62.5049&-18.2605&0&1.13798\\
6(a) & Y &2.1&1&-0.187165&-0.875099&-20.9172&99.1027&-170.089&130.938&-38.0072&0&1.13219\\
6(b)& LJ  &0&1&0.412147&1.61894&-5.73846&10.0637&-9.48287&4.71324&-0.974014&0&0.775053\\
6(b)& LJ  &0.1&1&0.374952&0.774098&-3.02401&5.77044&-5.69195&2.90088&-0.607194&0&0.695483\\
6(b)& LJ  & 0.2&1&-0.0259831&0.448125&-2.49345&8.72886&-13.2268&9.67743&-2.72339&0&1.32478\\
6(b)& LJ  &0.3&1&-0.110418&0.500301&-8.71096&38.0178&-72.4731&72.5622&-37.2913&7.78349&1.08473\\
6(b)& LJ  & 0.4&1&-0.426825&0.267704&-21.7157&122.223&-279.301&326.179&-193.284&46.2408&0.917036\\
7(a)& Y  & 1.82 & 1&-0.10833&-1.40042&-7.99612&47.6637&-85.1519&66.386&-19.3495&0&1.13885 \\
7(a)& LJ  & 0.4&1&-0.426825&0.267704&-21.7157&122.223&-279.301&326.179&-193.284&46.2408&0.917036\\
7(a)&  G  & 2.5 &1& -0.202165&-1.65283&-17.9331&135.977&-352.077&448.15&-283.891&71.6529&0.983802\\
7(a)& E   &1.813&1& -1.09031&-2.16095&30.576&-99.8449&165.53&-151.694&73.1886&-14.5204&0.823677\\
9& E   &1&0.1&0.422319&1.45854&-4.35437&3.93238&5.11002&-14.2398&11.6767&-3.39351&0.77295\\
9& E  &1&0.3&0.411692&1.11381&-2.86337&0.299185&10.7624&-19.7299&14.683&-4.09718&0.765508\\
9& E  &1&0.5&0.274977&0.375024&2.8439&-17.8841&43.0349&-52.6678&32.7557&-8.2269&0.788426\\
9& E  &1&0.7&0.0666694&-0.635408&10.4577&-39.9909&78.0552&-84.3057&48.0902&-11.3238&0.855997\\
9& E  &1&0.9&-0.0977131&-1.83569&16.2998&-50.1598&83.5545&-79.3539&40.6478&-8.74109&0.943813\\
9& E  &1&1.1&-0.683713&0.654502&11.6287&-50.4985&99.458&-105.867&59.1351&-13.6179&0.836162\\
9& E  &1&1.3&-0.868291&-0.0780018&10.9717&-29.3915&32.1351&-11.0863&-4.90455&3.32424&0.932054\\
9& E  &1&1.5&-1.0713&0.527968&-0.0153676&26.0912&-95.6713&142.44&-98.7294&26.4225&1.01572\\
9& E  &1&1.8&-1.43689&3.52772&-29.5133&143.46&-330.821&397.535&-242.33&59.4076&1.12934\\
9& E  &1&2.1&-1.86145&8.06742&-64.7503&260.356&-529.595&580.619&-329.22&76.0464&1.23315\\
[1ex] 
\hline 
\end{tabular}
\label{table:nonlin} 
\end{table*}


%

\end{document}